# Cloud Migration Methodologies: Preliminary Findings

Mahdi Fahmideh, Farhad Daneshgar, Fethi Rabhi

University of New South Wales, Sydney, Australia

**Abstract.** Research around cloud computing has largely been dedicated to addressing technical aspects associated with utilizing cloud services, surveying critical success factors for the cloud adoption, and opinions about its impact on IT functions. Nevertheless, the aspect of process models for the cloud migration has been slow in pace. Several methodologies have been proposed by both academia and industry for moving legacy applications to the cloud. This paper presents a criteria-based appraisal of such existing methodologies. The results of the analysis highlight the strengths and weaknesses of these methodologies and can be used by cloud service consumers for comparing and selecting the most appropriate ones that fit specific migration scenarios. The paper also suggests research opportunities to improve the status quo.

## 1       Introduction

Cloud computing initiatives have received significant attention for addressing computational requirements of enterprise applications through offering a wide range of services which are universally accessible, acquirable and releasable on the fly, and payable based on the service usage. Many IT-based organizations are at the edge of moving their legacies to the cloud. While there are many valuable technical solutions to make legacies cloud-enabled, those solutions are not sufficient on their own and one should not undermine the equal importance of adopting a systematic methodology to enable legacies to benefits from cloud services. Such a methodology aids developers to organize the migration process and defines a step-by-step guidance on activities should be carried out to reengineer and move legacies to the cloud. This paper presents a review and evaluation of the extant cloud migration methodologies in the literature with the aim of understanding their features, strengths, weaknesses, and potential opportunities for future research. More than a dozen of cloud migration methodologies have been suggested by both from academia and industry. Some examples are Chauhan's Methodology [1], REMICS [2], Tran's Methodology [3], Cloud-RMM [4], Strauch's Methodology [5], Zhang's Methodology [6], Oracle [7], ARTIST [8], Amazon [9], and Legacy-to-Cloud Migration Horseshoe [10].

This paper is organized as follows: Section 2 develops an evaluation framework to assess the abovementioned methodologies, which is followed by Section 3 that reports evaluation results of the methodologies. Section 4 discusses the implications and threats of this research. Finally, this paper concludes in Section 5.

## 2 Criteria-Based Evaluation of Migration Methodologies

In the context of software engineering, an evaluation framework constitutes a checklist of criteria (or methodological requirements) that an ideal methodology is expected to appropriately address them when dealing with a particular activity or objective [11]. A methodology is checked against an evaluation framework in two steps: firstly, the methodology is scanned for features concerned by a criterion and then an evaluation result (e.g. scale point) is yield signifying the some extent that the methodology supports the criterion [11]. In order to ensure the quality of criteria set, the meta-criteria (criteria used to assess other criteria) proposed by Karam et. al. [12] was taken into account to develop the proposed evaluation framework. Regarding to this source and context of this study, the criteria should be (i) sufficiently generic to cover a variety of scenarios for legacy application migration to the cloud regardless of a particular target cloud platform, (ii) distinct to characterise the similarities and differences of methodologies, and (iii) adequately comprehensive to cover end-to-end cloud migration process.

In developing of the criteria, we reviewed and synthesized various existing frameworks that define criteria attuned to evaluate software development methodologies in traditional software (re-engineering) engineering and software process improvement literatures. A set of criteria were identified in the studies by [13], [14], [15], [16], and [17]. We also reviewed studies suggesting criteria pertinent to cloud migration methodologies. Studies by [5], [3], [18], and [19] proposed essential criteria that an ideal cloud migration methodology should satisfy. This includes interoperability/portability, incompatibility resolution, cloud provider selection, and re-architecting, and tailorability. Once the criteria in the above sources were analyzed and redundancy and overlapping among them were removed, nineteen distinct criteria were shortlisted for the purpose of the current study. The criteria help to contrast and compare existing methodologies. They are listed in Table 1 and described in Section 3. We do not claim that the proposed evaluation framework is comprehensive, however, such a framework has not been proposed yet in literature and our study provides a good starting point in assessing and comparing extant cloud migration methodologies to highlight their strengths and weaknesses.

**Table 1.** Criteria expected to be supported by cloud migration methodologies

| Criterion and Definition (letter C is the unique identifier of each criterion) | Type |
|---|---|
| **Tailorability (C1)**: Providing mechanisms to configure and modify process or modelling language for a given project at hand. | Scale |
| **Development Roles (C2)**: Defining roles who are responsible for performing migration activities or any stakeholder who are involved in a migration process. | Scale |
| **Requirement Analysis (C3)**: Eliciting and specifying functionalities required to be fulfilled | Scale |

| | | |
|---|---|---|
| | by cloud-enabled application such as computational, security, elasticity, and storage space requirements. | |
| | **Legacy Understanding (C4)**: Recapturing an abstract As-Is representation of application architecture in terms of functionality, different types of dependencies to other applications, interaction points and message follows between application components, as well as quality of code blocks for reuse and adaptation. | Scale |
| | **Cloud Service Selection (C5)**: Identifying, evaluating, and selecting a set of cloud providers that might suit organization and application requirements. | Scale |
| Re-Architecting | **Cloud Architecture Model Definition (C6)**: Identifying components of legacy that are suitable for migration and defining their deployment in the cloud environments. | Scale |
| | **Refactoring and Incompatibility Resolution (C7)**: Identifying and resolving incompatibilities between legacy components and cloud services. | Scale |
| | **Enabling Application Elasticity (C8)**: Providing support for dynamic acquisition and release of cloud resources. | Scale |
| | **Enabling Multi-Tenancy (C9):** Providing support for enabling multi-tenancy in the application in terms of security, performance, customizability, and fault isolation, which might incur by running application in the cloud. | Scale |
| | **Deployment (C10)**: Adjusting the application and network configuration for the target cloud environment. | Scale |
| | **Monitoring (C11)**: Continuous monitoring of application and cloud resources to assure SLAs. | Scale |
| | **Test (C12)**: Defining activities for test and continuous delivery. | Scale |
| | **Work-Products and Notations (C13)**: Specifying work-products to be produced as outcomes of migration activities. | Scale |
| | **Modelling Language (C14)**: Specifying a modelling or notational component | Boolean |
| | **Unit of Migration (C15)**: Applicability of the methodology for the migrating different tier of a legacy application. | Multiple Answer |
| | **Migration Type (C16)**: Migration types are concerned with methodology. | Multiple Answer |
| | **Tool Support (C17)**: Availability of tools to support the methodology's activities and techniques. | Scale |
| | **Maturity (C18)**: Available account on successful adoption the methodology in real-world migration scenarios. | Multiple Answer |

## 3  Analysis of Results

Table 2 summarizes the evaluation results of the methodologies according to the framework. The review of the methodologies reveals that many criteria; specifically Tailorability (C1), Development Roles (C2), Cloud Architecture Definition (C6), Refactoring (C7), and Multi-tenancy (C9) are not adequately supported. The methodologies do not comprehensively elaborate on activities related to these criteria, as a part of their mainstream process, which should be carried out to make a legacy application cloud-enabled. The following, delineates the results of our analysis and suggests areas that indicate future research directions to improve existing methodologies.

**Tailorability (C1)**. It has been well-acknowledged that in every methodology there are good features to adopt as well as deficiencies to avoid [15]. These features circumscribe the applicability of a methodology in a given project situation at hand. The fact that methodologies should be tailored or designed to suit the characteristics of a given cloud migration scenario is pinpointed in the cloud migration literature [19, 20]. As shown in Table 2, except for REMICS and ARTIST methodologies which provide a partial support for the tailorability, none of the existing methodologies offers mechanisms to fine-tune their processes or to check if the methodology is properly applica-ble for a migration scenario at hand. REMICS is structured in the form of a set of reusable method fragments which eases its tailoring through selecting suitable method fragments and assembling them with respect to a migration scenario. ARTIST offers a tool which facilitates configuration and instantiation of the methodology for a given migration scenario. However, none of them provides explicit guidance on how to tailor or create a situational methodology out of the base methodology.

**Development Roles (C2)**. While methodologies describe what activities are to carry out, the roles and required expertise that take these activities become a concern for its users. The definition of roles assists developers who have limited experience and are not sure about roles involving in a migration process. In spite of necessity of defining roles in any software development lifecycle, the majority of cloud migration methodologies do not specify roles involving during the migration process. As shown in Table 2, except for Chauhan's methodology and ARTIST, the definition of roles and their responsibilities have been neglected in the existing methodologies. In these methodologies, the definition of roles is borrowed from traditional software development and they do not define roles that might be cloud-specific.

**Requirement Analysis (C3)**. Requirements specify the desired features that should be fulfilled by moving legacies to the cloud. Conventional requirement analysis techniques such as interview, prototyping, and workshop are widely used by REMICS, ARTIST, and Chauhan's methodology. Additionally, Oracle and Amazon extend the requirement analysis to focusing on computing requirements and application scalability. Furthermore, Legacy-to-Cloud Migration Horseshoe is concerned with interoperability requirements of the target application between cloud platforms. Trans and Zhang's methodologies do not define any activities related to the requirement analysis.

**Legacy Application Understanding (C4)**. This is common that the knowledge about legacy applications is undocumented and incomplete. An in-depth understanding of the current state of legacies helps to identify any characteristic that might influence the cloud migration process. Activities related to the legacy understanding are covered by most of the methodologies, except for Tran's and Oracle's methodologies. A few of reviewed methodologies such as Chauhan's methodology, Cloud-RMM, Strauch's methodology define activities related to recover of legacy architecture model but do not narrow to provide adequate mechanisms or guidelines to conduct them.

**Cloud Platform/Service Selection (C5)**. Developer should not neglect the influences of selecting cloud platforms on the development effort and cost required for the migration process. A better compatibility between the legacy and cloud services can make the migration process very easy and shorter. For example, Tran's methodology reports a breakdown of activities for moving a .NET 3-tier application to run in Windows Azure. She highlights required development efforts for modifying data tier, code refactoring, and installation is major if the underlying technology of the legacy and cloud platform are not compatible with each other. Table 2 shows that all of the reviewed methodologies incorporate activities related to cloud service selection. However, REMICS and Oracle are at the other end of the spectrum: they do not provide any guidelines as to how cloud service can be evaluated and selected.

**Re-Architecting (C6,C7,C8,C9)**. Several important aspects should be incorporated into the migration process when re-engineering a legacy to a cloud platform. The first (C6) is to select suitable legacy components and define their new deployment model in the cloud on basis of concerns such as network latency, data transfer, data privacy, legal restrictions while satisfying the expected QoS of the whole application. Cloud migration methodologies can be examined with respect to their support for activities and guidelines to define a cloud architecture model of an application and move components to the different cloud servers. Only Chauhan's methodology defines this activity in its process model. The second architectural aspect (C7) is the resolving of incompatibilities that might occur between the legacy application and selected cloud platform/services. The incompatibilities might be sourced from mismatch between legacy codes and cloud service APIs, interface signatures, data types, and query calls. A methodology is expected to provide activities to identify possible incompatibility issues and accordingly proper guidelines to resolve them. Otherwise legacy will not be able to utilize cloud services. Back to Table 2, the criteria refactoring and incompatibility resolution is only supported by Strauch's methodology and Oracle methodology, though they focus on activities to resolve incompatibilities between the legacy database tier and a target cloud database service. Other methodologies suggested by Chauhan, REMICS, Tran, Cloud-RMM, ARTIST, and Amazon suffer from cursory definitions of code refactoring. The third architectural design aspect (C8) is to enable the legacy in a support of dynamic resource acquisition and release when it is running in the cloud. According to Table 2, activities related to the enabling elasticity in the legacies are only supported by Amazon methodology. The fourth aspect (C9) is the multi-tenancy support. A key concern in the re-architecting of legacy to address multi-tenancy is to provide a support in the application for isolating the security, performance, customizability, and fault of tenants. Without such a support, a migrated application may face the risk of tenant interference. As shown in Table 2, the majority of methodologies are silent regarding the multi-tenancy requirement. Cloud-RMM includes the multitenancy, however, it does not elaborate on how conducing it.

**Deploying and monitoring (C10, C11)**. It is likely that the connection between the migrated legacy to the cloud and local network to be required. A methodology should properly define activities related to the network configuration such as the setting open ports, firewall policies, and connection strings to data and application (C10). The deployment is covered by all the methodologies except for Chauhan's methodology and Legacy-to-Cloud Migration Horseshoe. Besides, once migrated to the cloud, the continuous monitoring of the application and cloud resources to assure SLAs is necessary (C11). Only three methodologies Oracle, ARTIST, and, Amazon support this criterion.

**Test (C12)**. Test is to ensure that the cloud-enabled application meets the goals of cloud migration. All the methodologies except for Strauch's methodology, Zhang's methodology, and Legacy-to-Cloud Migration Horseshoe define activities in coherence with the methodology to ensure that application conforms to the expectation of the cloud migration such as performance or resource utilization.

**Work-products and modelling language (C13 and C14).** An integral part of every methodology is to specify necessary work-products as the outcome of each activity throughout the process model. Defining work-product becomes important if automatic code generation is required for a specific cloud platform or users of methodology aim to trace or keep the list of work-products that have been produced throughout the migration process. With respect to this, among the reviewed methodologies, Chauhan's methodology and ARTIST explicitly define work-products as a result of performing each migration activity. However, REMICS, Zhang's methodology, and Legacy-to-Cloud Migration Horseshoe only defines representing the legacy architecture.

Methodologies may specify modelling techniques along with a particular notation to represent the outcome of each development activity. Modelling techniques, however, is hardly supported in the existing methodologies. ARTIST and REMICS use UML for their whole lifecycles along with Zhang's methodology and Legacy-to-Cloud Migration Horseshoe that, respectively, use SoaML and Graph-based modelling to represent legacy architecture.

**Unit of Migration (C15)**. Some organizations may not move the whole legacy stack to the cloud because of security concerns, rather they may migrate some legacy components to the cloud whilst other components are kept in local organization network and cloud services are offered to them. In this regard, it is important to investigate if a methodology is appropriate for moving a particular tier or whole legacy stack to the cloud. Given that, eight reviewed methodologies have been designed for full migration to the cloud. Strauch's methodology is particularly designed for moving legacy data tier to a cloud database solution.

**Migration Type (C16)**. Regarding the common service delivery models IaaS, SaaS, and PaaS, one can view that there are several variants that a legacy can utilize cloud services. We defined the followings migration variants and assess if the methodologies support them. In *Type I* the business logic tier of a legacy (e.g. WS-BPEL), which offers discrete and reusable functionality, is deployed in the cloud infrastructure. In this migration type, the data tier is kept in local organization network. Deploying an image processing component of a legacy in E2C is an example of this migration type. In *Type II* some components or whole application stack is replaced with an available and fully tested cloud service. The Salesforce CRM application is a typical example of this type of cloud migration. In this review, Chauhan's methodology, REMICS, Zhang's methodology, and ARTIST support this type of migration. In *Type III* legacy database is deployed in a cloud data store provider. The components related to business logic tier are kept in local organization network and the database is deployed in public cloud data store such as Amazon Simple Storage Service, Amazon Elastic Block Store, Dropbox, or Zip Cloud. There is not methodology to support this migration type. In *Type IV* the data tier of a legacy is modified and converted to a cloud database solution such as Amazon SimpleDB, Google App Engine data store, or Google Cloud SQL. Tran's, Strauch's methodology, and Oracle support this migration type. Finally, in *Type V* the whole legacy stack is deployed in the cloud infrastructure where the legacy is encapsulated in a single virtual machine and then run in the cloud infrastructure. From the reviewed methodologies, Amazon defines activities to carry out this migration type. Obviously, on the basis of a chosen migration type, different activities might be required to be carried out and accordingly a methodology should properly address them.

**Tool support (C17).** The adoption of a methodology is facilitated if it offers its own supportive tool or alternatively refers developers to existing third-party tools available in the cloud marketplace. Only ARTIST and REMICS provide tool for whole migration process model. More specifically, ARTIST proposes Eclipse-based suites which are integrated with its activities. Since produced work-products are stored in a shared repository, they can be accessed and modified by other tools. REMICS includes a set of tools that can be classified in the areas such as requirement management, knowledge recovery from legacies, re-transformation of legacy components to cloud architecture, and model-based testing. On the other hand, Strauch's and Amazon's methodologies offer tool for legacy architecture recovering, data migration, and resource elasticity management. Other methodologies do not offer any tools.

**Maturity (C18).** Validating a methodology in real-world migration scenarios and subsequently refining it through feedback from experts improves its applicability and maturity. In this review, the majority of methodologies, except for Cloud-RMM and Legacy-to-Cloud Migration Horseshoe, have reported the applying the methodology in a real-world example. An observed issue during the assessment was the lack of sufficient contextual information on the environment in which the methodology had been applied, description of techniques used to data collection and analysis, and addressing threats to validity.

**Table 2.** Results of evaluating cloud migration methodologies

| Criterion | Chauhan's Methodology | REMICS | Tran's Methodology | Cloud-RMM | Strauch's Methodology | Zhang's Methodology | Oracle | ARTIST | Amazon | Legacy-to-Cloud Migration Horseshoe |
|---|---|---|---|---|---|---|---|---|---|---|
| C1  | ○ | ◐ | ○ | ○ | ○ | ○ | ○ | ◐ | ○ | ○ |
| C2  | ● | ◐ | ○ | ○ | ○ | ○ | ○ | ● | ○ | ○ |
| C3  | ● | ● | ○ | ◐ | ◐ | ○ | ● | ● | ● | ● |
| C4  | ◐ | ● | ○ | ◐ | ◐ | ● | ○ | ● | ● | ● |
| C5  | ● | ○ | ● | ◐ | ● | ● | ○ | ● | ● | ● |
| C6  | ● | ◐ | ○ | ◐ | ○ | ○ | ○ | ○ | ◐ | ○ |
| C7  | ◐ | ◐ | ◐ | ◐ | ● | ● | ○ | ◐ | ◐ | ○ |
| C8  | ○ | ○ | ○ | ◐ | ○ | ○ | ○ | ○ | ● | ○ |
| C9  | ○ | ○ | ○ | ◐ | ○ | ○ | ○ | ○ | ○ | ○ |
| C10 | ○ | ◐ | ● | ◐ | ◐ | ● | ● | ● | ● | ○ |
| C11 | ○ | ○ | ○ | ○ | ○ | ○ | ● | ● | ● | ○ |
| C12 | ◐ | ● | ● | ◐ | ○ | ○ | ● | ● | ● | ○ |
| C13 | ● | ● | ○ | ○ | ○ | ◐ | ○ | ● | ○ | ● |
| C14 | N | Y | N | N | N | Y | N | Y | N | Y |
| C15 | AS | AS | AS | NS | DL | AS | AS | AS | AS | AS |
| C16 | II | II | IV | NS | IV | II | IV | II | V | NS |
| C17 | ○ | ● | ○ | ○ | ◐ | ○ | ○ | ● | ◐ | ○ |
| C18 | RE | CS | RE | NV | CS | CS | RE | CS | CS | NV |

●*Fully Supported* explicitly supported by the method, ◐*Partially Supported* by the method, ○*Not Supported* by the method (neither a partial definition nor explanation for a requirement), N:No, Y:Yes, NV: Not Validated, CS: Case Study, RE: Reported Experience, AS: Application Stack, DL: Data Tier, NS: Not Specified, I, II, III, IV, and V: See definitions for the requirement Migration Type in Section 4.

# 4     Discussion

This section discusses research implications and possible threats to validity of the evaluation results.

**Research implications**. This research has two major contributions to the cloud migration literature. Firstly, the proposed evaluation framework serves as a valuable tool for project managers to assess and compare the capabilities of cloud migration methodologies and select ones which satisfies their migration scenario characteristics through reusing the strengths. They can also priorities the proposed requirements on the basis of their goals and evaluate methodologies with respect to these priorities. Secondly, the evaluation results can be used as a basis for the purpose of situational cloud migration methodology construction meaning that useful method fragments from the existing methodologies can be selected and assembled to create bespoke methodology that fits the characteristics of a migration scenario at hand.

**Threats to evaluation validity**. In spite of our effort to provide a comprehensive and objective comparison, some threats still exist as mentioned in the followings:

*Conclusion validity*. The evaluation results in this research are mainly theoretical and based on the available and published documents of the methodologies. However, a real evaluation of the methodologies through applying them in the same real-world migration scenario could yield to other results. However, such empirical assessment is planned as our future work. Furthermore, as the methodologies may have been yet improved by their designers, the evaluation may need to be updated.

*Construct validity*. The validity of the evaluation results may be concerned in terms of measures that have been applied to assess the satisfaction of the methodological requirements. To minimize inconsistency in measuring, the definitions of criteria were used during assessment process (Table 1). These definitions checked the existence of activities, work-products, or roles that are related to the criteria.

*Internal validity*. A threat to the validity of this research is that the evaluation process was conducted by a single researcher. Hence, the evaluation results might be to some extent subjective in nature or undergone by misinterpretation. This threat can be further minimized if a Delphi technique [21] is applied where the evaluation process is performed by authors of the framework and subject matter experts. The difference between ratings can be resolved through a post-hoc evaluation discussion to reach a consensus.

*External validity*. We acknowledge that to assure generalizability of evaluation results, more evaluation with a higher number of domain experts is necessary. Furthermore, we selected a representative number of cloud migration methodologies that have been proposed in the literature. Nevertheless, more research on the evaluation of cloud migration methodologies and criteria which may have not been investigated in this research is required.

## 5      Future Work and Conclusion

This paper argued that the current state of legacies to the cloud needs to adopt methodological/process model perspective. Following an overview of existing methodologies related to legacy application to cloud migration, an evaluation of them regarding a set of important criteria was presented. The evaluation results were presented in a structured format and revealed that the methodologies suffer from the lack of tailorability, defining roles and work-products involved in the migration process, and incorporating a modelling language to model the output of activities. Additionally, there is no methodology which focuses on the migration types I and III. The cloud migration methodologies seem quite nascent and are yet to reach a high level of maturity. The current situation of the cloud migration methodologies definitely calls for further research aimed at ameliorating the status quo. With respect to the evaluation results in this paper, a further research opportunity is to develop a new cloud migration methodology through reusing the strengths of existing methodologies while addressing their deficiencies. This can be based on identifying method fragments from the methodologies and storing them in a method library. Such a harness can be effectively addressed by adopting the idea of *situational method engineering* approach [22] where cloud migration solutions in the literature can be abstracted away and structured as a complementary source for developing method fragments. Once such reusable method fragments identified, they can be further assembled to construct customized migration methodology which fits a given migration scenario.